\documentclass[aps,pra,twocolumn,showpacs]{revtex4}
\usepackage{amssymb}
\usepackage{amsmath}
\usepackage{epsfig}

\begin{document}

\title{Coupled-mode theory for binary optical lattices}

\author{Lasha Tkeshelashvili}
\affiliation {
Andronikashvili Institute of Physics, Tbilisi State University, 
Tamarashvili 6, 0177 Tbilisi, Georgia}

\begin{abstract}
The coupled-mode theory is developed for description of the nonlinear wave dynamics in binary optical lattices. The obtained equations of motion accurately describe nonlinear wave dynamics close to the band edges and in the gap of the linear spectrum of the system. In order to demonstrate the power of the presented approach, bright gap solitary wave solutions of the nonlinear coupled-mode equations are derived and examined both analytically and numerically. The presented results are relevant to nonlinear wave phenomena in coupled waveguide arrays, coupled nano-cavities in photonic crystals, metallo-dielectric systems, and the  Bose-Einstein condensates in deep optical lattices.
\end{abstract}

%42.65.Tg Optical solitons; Nonlinear guided waves
%78.67.Pt optical properties of superlattices
%42.70.Qs Photonic band gap materials

\maketitle

Periodically modulated nanostructures may exhibit band gaps in their linear spectrum \cite{joannopoulos2008}. That gives rise to  a number of unique optical properties of those artificial systems. A prominent example is the existence of so-called gap solitons \cite{chen1987}. Solitons are localized nonlinear wave packets that can propagate undistorted in dispersive media \cite{ablowitz2011}. The gap solitons have carrier wave frequency in the stop gap where the propagation of linear modes is not allowed. A powerful framework for description of nonlinear wave phenomena in photonic band gap materials is proven to be offered by the coupled-mode approach \cite{desterke1994, busch2007, desterke1996}. Indeed, various types of solitary wave solutions of the nonlinear coupled-mode equations, including bright, dark, and anti-dark gap solitons were found recently \cite{iizuka2000, alatas2003, alatas2006, alatas2007}.

Perhaps, coupled optical waveguide arrays represent the most convenient experimentally accessible systems for studies of linear and nonlinear wave dynamics \cite{kivshar2003}. That includes the realization of optical analogies of various quantum-mechanical phenomena \cite{longhi2009}, as well as the demonstration of effects related to the discrete optical solitons \cite{lederer2008}. A theoretical model that provides quantitative description of the waveguide arrays is the discrete nonlinear Schr\"odinger (DNLS) equation which, in the dimensionless form, can be written as follows~\cite{kevrekidis2009}:
\begin{equation}
i \frac{\partial \psi_n}{\partial z} + C (\psi_{n+1} + \psi_{n-1}) + \varepsilon_n \psi_n + N |\psi_n|^2 \psi_n  = 0 \, .
\label{dnlse}
\end{equation}
It is worth to mention here that DNLS equation is used for studies of coupled nano-cavities in photonic crystals \cite{christodoulides2002}, metallo-dielectric systems \cite{ye2010}, and the  Bose-Einstein condensates in deep optical lattices \cite{morsch2006} too. In the context of the waveguide arrays, in Eq.~\eqref{dnlse}, $z$ represents the propagation coordinate along the waveguides, $\psi_n$ is the eigenmode amplitude in the $n$-th waveguide, and $C$ gives the coupling rate between adjacent sites. In what follows it is taken to be a positive constant. $N$ is the effective nonlinear constant of the system. The $n$-dependent term $\varepsilon_n$ is caused by the different effective refractive index of the individual waveguides.

Binary waveguide arrays consist of the sites with alternating effective refractive index. That is, $\varepsilon_n = \varepsilon_{\mathrm{a}}$ for sites with $n = 0, \pm 1, \cdots$, and $\varepsilon_n = \varepsilon_{\mathrm{b}}$ for sites with $n = \pm 2, \cdots$. Before proceeding further, it should be noted that only the variation in $\varepsilon_n$ is relevant for the wave dynamics. Indeed, by means of the gauge transformation $\psi_n \rightarrow \exp(i\tilde{\varepsilon} z)\psi_n$, all $\varepsilon_n$ can be shifted by arbitrary $\tilde{\varepsilon}$. The choice $\tilde{\varepsilon} = (\varepsilon_{\mathrm{a}} + \varepsilon_{\mathrm{b}})/2$ leads to $\varepsilon_n = (-1)^n \varepsilon$ with $\varepsilon = (\varepsilon_{\mathrm{a}} - \varepsilon_{\mathrm{b}})/2$. Thus, without loss of generality, for the modulation term we write
\begin{equation}
\varepsilon_n = \varepsilon \exp(\pm i\pi n) =\varepsilon \cos(2k_0n) \, ,
\label{modulation}
\end{equation}
where $k_0=\pi/2$, and $\varepsilon$ is assumed to be positive. This recovers the fact that, Eqs.~\eqref{dnlse} and \eqref{modulation} describe periodically modulated systems with the spatial period $\pi / k_0 = 2$. 

To obtain the linear dispersion relation for the considered system let us write the linearized DNLS equation for the even- and odd-numbered sites separately
\begin{align}
& i \frac{\partial \psi_{2n} }{\partial z} + C (\psi_{2n+1} + \psi_{2n-1}) + \varepsilon \psi_{2n} = 0 \, , \nonumber \\
& i \frac{\partial \psi_{2n+1} }{\partial z} + C (\psi_{2n+2} + \psi_{2n}) + \varepsilon \psi_{2n+1}  = 0 \, .
\label{linear}
\end{align}
Then, inserting the following ansatz \cite{kittel1976}:
\begin{align}
& \psi_{2n}  = \Psi_{\mathrm{e}} \exp(-i[\omega z - 2n k]) \, , \nonumber \\
& \psi_{2n+1}  =  \Psi_{\mathrm{o}} \exp(-i[\omega z - (2n+1) k]) \, ,
\label{linearansatz}
\end{align}
with some constants $\Psi_{\mathrm{e}}$ and $\Psi_{\mathrm{o}}$, it is straightforward to  derive the linear dispersion relation 
\begin{equation}
\omega = \pm \sqrt{\varepsilon^2 +4 C^2 \cos^2(k)} \, .
\label{dispersion}
\end{equation}
Thus, the linear spectrum of the system consists of two bands separated by a band gap which ranges from $-\varepsilon$ to $+\varepsilon$. As expected, for $\varepsilon = 0$ Eq.~\eqref{dispersion} reduces to the dispersion relation of the uniform system  $ \omega = \pm 2 C \cos(k)$. In this expression the sign is related to the choice of the positive direction. In what follows we choose the minus sign whenever the modulation is neglected.

Recently, a number of exciting wave phenomena associated with the peculiar dispersive properties of the binary
DNLS equation were demonstrated in both linear \cite{longhi2011} and nonlinear \cite{sukhorukov2002, morandotti2004,conforti2011, conforti2012} regimes. In the present article the coupled-mode theory is developed for description of the nonlinear wave dynamics in binary waveguide arrays.

If $\varepsilon$ and $N$ were zero, a solution of Eq.~\eqref{dnlse} would consist of the forward and backward propagating plane waves
\begin{equation}
\psi_n = f_n \exp(-i[\omega_0 z - k_0n]) + b_n \exp(-i[\omega_0 z + k_0n]) \, ,
\label{cmeansatz}
\end{equation}
with constant amplitudes $f_n$ and $b_n$. Here, $\omega_0$ is given by Eq.~\eqref{dispersion} at $k = k_0$. Note that, since $\varepsilon = 0$ is assumed, $\omega_0 = 0$. That is due to the particular choice of the gauge, and therefore, has no relevance to the system dynamics.

The coupled-mode approach is based on the observation that the periodic modulation given by Eq.~\eqref{modulation} causes strong coupling between the forward and backward propagating waves \cite{desterke1994}. Indeed, taking into account that $k_0 = \pi/2$ and using Eqs.~\eqref{modulation} and \eqref{cmeansatz} we can write
\begin{align}
\varepsilon_n \psi_n = & \   \varepsilon  f_n \exp(-i[\omega_0 z - k_0 n] - i\pi n)\nonumber \\
                       & \qquad + \varepsilon b_n \exp(-i[\omega_0 z + k_0 n] +i\pi n) \nonumber \\
                     = & \   \varepsilon b_n \exp(-i[\omega_0 z - k_0 n])  \nonumber \\
                       & \qquad + \varepsilon f_n \exp(-i[\omega_0 z + k_0 n]) \, .
\label{lcoupling}
\end{align}
As a result, in the modulated systems, $f_n$ and $b_n$ become dependent on $n$. 

Besides that, $f_n$ and $b_n$ vary due to the nonlinear effects. By inserting Eq.~\eqref{cmeansatz} in $|\psi_n|^2\psi_n$, in the calculations appear so-called "non-phase-matched" terms involving $\exp(\pm i 3 k_0 n)$. In the optical context it is often argued that such effects are unimportant for the system dynamics \cite{desterke1994}. However, since $\exp(\pm i 3 k_0 n) = \exp(\mp i k_0 n)$ holds in the present case, those terms must be retained along with the phase-matched ones \cite{desterke1993}.

Now, assuming that the wave envelope is slowly varying with respect to the lattice spacing, $n$ can be treated as a continuous variable. That yields,
\begin{align}
& f_{n \pm 1} \approx f(z,n) \pm \frac{\partial f(z,n)}{\partial n} \, , \nonumber\\
& b_{n \pm 1} \approx b(z,n) \pm \frac{\partial b(z,n)}{\partial n} \, ,
\label{sveder}
\end{align}
where $f_{n}(z) \equiv f(z,n)$ and $ b_{n}(z) \equiv b(z,n)$. 

Finally, inserting Eq.~\eqref{cmeansatz} in Eq.~\eqref{dnlse} and collecting the terms with $\exp(+ik_0n)$ and $\exp(-ik_0n)$ respectively, it is straightforward to derive the following nonlinear coupled-mode equations \cite{desterke1993}:
\begin{align}
i\left( \frac{1}{v_{\mathrm{g}}} \frac{\partial f}{\partial z} +  \frac{\partial f}{\partial n} \right)  +\kappa b + \Gamma (|f|^2 + & 2  |b|^2 ) f  \nonumber \\
& +  \Gamma b^2 f^{\ast} = 0 \, , \nonumber \\
i\left( \frac{1}{v_{\mathrm{g}}} \frac{\partial b}{\partial z} -  \frac{\partial b}{\partial n} \right) +\kappa f + \Gamma (|b|^2  + & 2  |f|^2 ) b \nonumber \\
& +  \Gamma f^2 b^{\ast} = 0  \, ,
\label{gcme}
\end{align}
where $\kappa = \varepsilon/v_{\mathrm{g}}$, $\Gamma = N / v_{\mathrm{g}} $, and $v_{\mathrm{g}} \equiv 2C$ is the group velocity of waves with $k = k_0$ in the uniform system, i.e. for $\varepsilon = 0$.

In order to estimate accuracy of the involved approximations let us insert 
\begin{align}
& f(z,n) = F \exp(-i[\Omega z - K n ]) \, , \nonumber \\
& b(z,n) = B \exp(-i[\Omega z - K n]) \, ,
\label{lcmeansatz}
\end{align}
into the linearized Eq.~\eqref{gcme}. Here $F$ and $B$ are some constants. That results in the linear dispersion relation
\begin{equation}
\Omega = \pm \sqrt{\varepsilon^2 +4 C^2 K^2} \, .
\label{lcmedispersion}
\end{equation}
By assuming $k = k_0 + K$ in Eq.~\eqref{dispersion}, it is easy to see that Eq.~\eqref{lcmedispersion} accurately reproduces the
system linear spectrum for $|K| \ll \pi/2$. Moreover, as is noted above, in Eq.~\eqref{gcme} the nonlinear response of the system is evaluated exactly. Therefore, Eq.~\eqref{gcme} is a valid model for studies of nonlinear wave dynamics in the band gap and/or close to the band edges \cite{desterke1994, busch2007}.

In order to demonstrate the power of the presented approach, let us consider the gap soliton solutions. The nonlinear coupled-mode equations exhibit rich spectrum of various nonlinear excitations including bright, dark, and anti-dark gap solitary waves \cite{iizuka2000, alatas2003, alatas2006, alatas2007}. 

Here, as an example, we consider bright gap solitons. In particular, following Ref.~\cite{iizuka2000}, it is straightforward to show that so-called "in-gap" traveling localized solutions of Eq.~\eqref{gcme} read
\begin{align}
& f(z,n) = \Delta^{-1/2} G(\zeta) \exp(i\theta_{\mathrm{f}}) \, , \nonumber \\
& b(z,n) = \Delta^{+1/2} G(\zeta) \exp(i\theta_{\mathrm{b}}) \, ,
\label{gsansatz}
\end{align}
where $\zeta = n -v z$. Moreover, $\Delta$ is a real constant defined as follows
\begin{equation}
\Delta = \sqrt{ \frac{v_{\mathrm{g}} - v}{v_{\mathrm{g}} + v}} \, ,
\label{delta}
\end{equation}
and so, $|v/v_{\mathrm{g}}|\le 1$. It is clear that $v$ parametrizes the soliton propagation velocity. The pulse amplitude reads 
\begin{equation}
G(\zeta) = \left[ \frac{\pm 2 \kappa \beta}{\Gamma \cosh(\xi)(1+\beta\cos(2\phi))} \right]^{1/2} \, ,
\label{gfactor}
\end{equation}
with
\begin{equation}
\phi = -2 \arctan([\tanh(\xi/2)]^{\mp 1} ) \, ,
\label{phi}
\end{equation}
and $\xi = 2 \kappa \gamma (\zeta - \zeta_0)$, where $\zeta_0$ is an arbitrary constant. In Eq.~\eqref{gfactor} the choice of the sign must guarantee $G(\zeta)$ to be a real function. Furthermore,
\begin{equation}
\gamma = \frac{1}{\sqrt{1 - (v/v_{\mathrm{g}})^2}} \, ,
\label{gamma}
\end{equation}
is the Lorentz factor, and
\begin{equation}
\beta = \frac{1}{1+2\gamma^2} \, .
\label{beta}
\end{equation}
Finally, the phases of the forward and backward propagating waves are given by
\begin{align}
& \theta_{\mathrm{f}} = \frac{2\beta\gamma^2}{\sqrt{1-\beta^2}}\frac{v}{v_{\mathrm{g}}} \arctan\left(\sqrt{\frac{1-\beta}{1+\beta}} \tan(\phi)\right) + \frac{1}{2}\phi \, , \nonumber \\
& \theta_{\mathrm{b}} = \frac{2\beta\gamma^2}{\sqrt{1-\beta^2}}\frac{v}{v_{\mathrm{g}}} \arctan\left(\sqrt{\frac{1-\beta}{1+\beta}} \tan(\phi)\right) - \frac{1}{2}\phi \, .
\label{phases}
\end{align}
It should be noted that the derived expressions represent a limiting case of more general two-parameter solitary wave solution \cite{iizuka2000}.

\begin{figure}[t]
{\epsfig{file=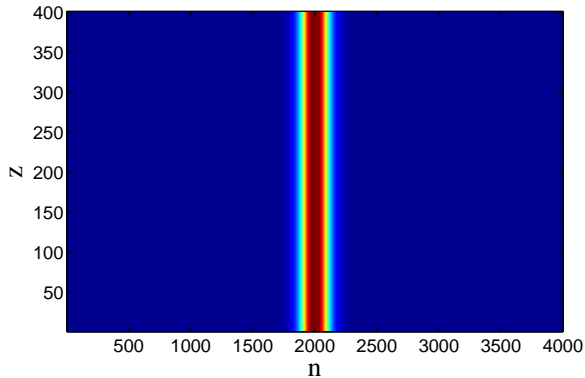,width=8.5cm}} 
\caption{(Color online) Sum of the envelope intensities $|f_n|^2 + |b_n|^2$ for the stationary, i.e. moving with $v = 0$, gap solitary wave. The presented results are obtained from the corresponding numerical solution of the DNLS equation. The chosen parameters are $C = 1$, $N = -1$, and $\varepsilon = 0.02$. The plotted variables are dimensionless.} 
\label{fig1}
\end{figure}

\begin{figure}[t]
{\epsfig{file=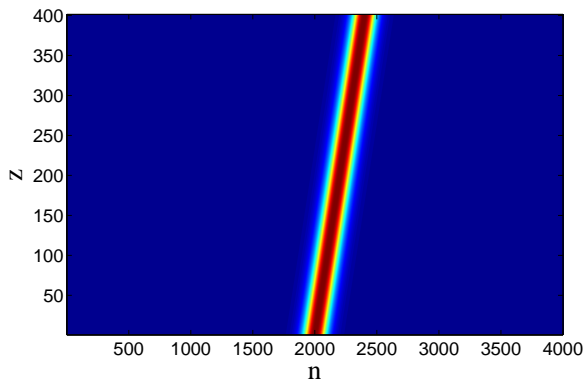,width=8.5cm}} 
\caption{(Color online) Sum of the envelope intensities $|f_n|^2 + |b_n|^2$ for the gap soliton moving with $v=v_{\mathrm{g}}/2$ velocity. The results are obtained from the corresponding numerical solution of the DNLS equation. The chosen parameters are $C = 1$, $N = -1$, and $\varepsilon = 0.02$. Note that $v=1$ in this case. The plotted variables are dimensionless.} 
\label{fig2}
\end{figure}

In addition, to demonstrate stability of the presented gap soliton solutions, we solve numerically the DNLS equation. In particular, the initial value problem for Eq.~\eqref{dnlse} is defined by Eqs.~\eqref{cmeansatz} and \eqref{gsansatz} at $z=0$. The simulation results depicted in Figs.~\ref{fig1} and \ref{fig2} show that the stationary as well as the mobile gap solitons are stable and propagate undistorted in the system. 

%%%%%%%%%%%%%%%%%%%%%%%%%%%%%%%%%%%%%%%%%%%%%%%%%%%%%%%%%%%%%%%%%%%%%%%%%%%%
%%%%%%%%%%%%%%%%%%%%%%%%%%%%%%%%%%%%%%%%%%%%%%%%%%%%%%%%%%%%%%%%%%%%%%%%%%%%

%%%%%%%%%%%%%%%%%%%%%%%%%%%%%%%%%%%%%%%%%%%%%%%%%%%%%%%%%%%%%%%%%%%%%%%%%%%%
%%%%%%%%%%%%%%%%%%%%%%%%%%%%%%%%%%%%%%%%%%%%%%%%%%%%%%%%%%%%%%%%%%%%%%%%%%%%%

The suggested couple-mode theory provides opportunity to establish close similarity between nonlinear wave phenomena in binary waveguide arrays and other photonic band gap materials such as, for example, fiber Bragg gratings. 
This fact offers a number of obvious advantages. First of all, the presented approach immediately shows that various types of solitary waves including bright, dark, and anti-dark gap solitons  \cite{iizuka2000, alatas2003, alatas2006, alatas2007} exist in binary waveguide arrays. On the other hand, the coupled waveguide arrays represent one of the most convenient systems to observe those nonlinear excitations which are difficult, or even impossible to study experimentally in another optical context \cite{desterke1996}.

To conclude, in the present article the coupled-mode theory is developed and applied to describe nonlinear wave dynamics in binary waveguide arrays. It is shown that the suggested equations represent an accurate model for studies of nonlinear waves in the band gap and/or close to the band edges. As an example bright soliton solutions of the nonlinear coupled-mode equations are derived. The stability of the found solitary waves is studied numerically. In particular, the numerical simulations of the DNLS equation show that the gap solitons propagate undistorted over long distances. Finally, it should be stressed that the presented results are relevant to nonlinear wave phenomena in coupled nano-cavities in photonic crystals \cite{christodoulides2002}, metallo-dielectric systems \cite{ye2010}, and the  Bose-Einstein condensates in deep optical lattices \cite{morsch2006} as well.

\textit{Acknowledgments.} This work is supported by Georgian National Science Foundation (Grant No.~30/12).


\begin{thebibliography}{abc}

\bibitem{joannopoulos2008}
J.~D.~Joannopoulos, R.~D.~Meade, and J.~N.~Winn,
\textit{Photonic Crystals: Molding the Flow of Light}, 
2nd ed. (Princeton Univ. Press, Princeton, 2008).

\bibitem{chen1987}
W.~Chen and D.~L.~Mills,
Phys. Rev. Lett. \textbf{58}, 160 (1987).

\bibitem{ablowitz2011}
M.~J.~Ablowitz,
\textit{Nonlinear Dispersive Waves: Asymptotic Analysis and Solitons} 
(Cambridge Univ. Press, Cambridge, 2011).

\bibitem{desterke1994}
C.~M.~de~Sterke and J.~E.~Sipe,
Prog. Opt. \textbf{33}, 203 (1994).

\bibitem{busch2007}
K.~Busch \textit{et al.},
Phys. Rep. \textbf{444}, 101 (2007).

\bibitem{desterke1996}
C.~M.~de~Sterke, D.~G.~Salinas, and J.~E.~Sipe,
Phys. Rev. E \textbf{54}, 1969 (1996).

\bibitem{iizuka2000}
T.~Iizuka and C.~M.~de~Sterke,
Phys. Rev. E \textbf{61}, 4491 (2000).

\bibitem{alatas2003}
H.~Alatas, A.~A.~Iskandar, M.~O.~Tjia, and T.~P.~Valkering,
J. Nonlinear Opt. Phys. Mater. \textbf{12}, 157 (2003).

\bibitem{alatas2006}
H.~Alatas, A.~A.~Iskandar, M.~O.~Tjia, and T.~P.~Valkering,
Phys. Rev. E \textbf{73}, 066606 (2006).

\bibitem{alatas2007}
H.~Alatas,
Phys. Rev. A \textbf{76}, 023801 (2007).

\bibitem{kivshar2003}
Y.~S.~Kivshar and G.~P.~Agrawal,
\textit{Optical Solitons: From Fibers to Photonic Crystals} 
(Academic Press, San Diego, 2003).

\bibitem{longhi2009}
S.~Longhi,
Laser Photon. Rev. \textbf{3}, 243 (2009).

\bibitem{lederer2008}
F.~Lederer \textit{et al.},
Phys. Rep. \textbf{463}, 1 (2008).

\bibitem{kevrekidis2009}
P.~G.~Kevrekidis, 
\textit{The Discrete Nonlinear Schr\"odinger Equation: Mathematical Analysis, 
Numerical Computations, and Physical Perspectives} 
(Springer, Berlin, 2009).

\bibitem{christodoulides2002}
D.~N.~Christodoulides and N.~K.~Efremidis,
Opt. Lett. \textbf{27}, 568 (2002).

\bibitem{ye2010}
F.~Ye, D.~Mihalache, B.~Hu, and N.~C.~Panoiu,
Phys. Rev. Lett. \textbf{104}, 106802 (2010).

\bibitem{morsch2006}
O.~Morsch and M.~Oberthaler,
Rev. Mod. Phys. \textbf{78}, 179 (2006).

\bibitem{kittel1976}
C.~Kittel,
\textit{Introduction to Solid State Physics}, 
5th ed. (Wiley, New York, 1976).

\bibitem{longhi2011}
S.~Longhi,
Appl. Phys. B \textbf{104}, 453 (2011).

\bibitem{sukhorukov2002}
A.~A.~Sukhorukov and Y.~S.~Kivshar, 
Opt. Lett. \textbf{27}, 2112 (2002).

\bibitem{morandotti2004}
R.~Morandotti \textit{et al.}, 
Opt. Lett. \textbf{29}, 2890 (2004).

\bibitem{conforti2011}
M.~Conforti, C.~De~Angelis, and T.~R.~Akylas,
Phys. Rev. A \textbf{83}, 043822 (2011).

\bibitem{conforti2012}
M.~Conforti, C.~De~Angelis, T.~R.~Akylas, and A.~B.~Aceves,
Phys. Rev. A \textbf{85}, 063836 (2012).

\bibitem{desterke1993}
C.~M.~de~Sterke,
Phys. Rev. E \textbf{48}, 4136 (1993).

\end{thebibliography}
\end{document}